\documentclass[%
reprint,
superscriptaddress,
amsmath,amssymb,
aps,
prl,
]{revtex4-1}
\usepackage{mwe}
\usepackage{bm}
\usepackage{graphicx}
\usepackage[normalem]{ulem}
\usepackage{graphicx}
\usepackage{epstopdf}
\usepackage{hyperref}
\usepackage[normalem]{ulem}
\usepackage{xcolor}
\usepackage{bm}
\usepackage{url}
\usepackage{lipsum}

\usepackage{bm}

\newcommand{\SrBiSe}{$\mathrm{Sr_{x}Bi_{2}Se_{3}}$ }
\newcommand{\CuBiSe}{$\mathrm{Cu_{x}Bi_{2}Se_{3}}$ }
\newcommand{\BiSe}{$\mathrm{Bi_{2}Se_{3}}$}	

\begin{document}

\title{Link between superconductivity and a Lifshitz transition in intercalated \BiSe}

\author{A. Almoalem}
\email{avioral@tecnhion.ac.il}
\affiliation{Department of Physics, Technion -- Israel Institute of Technology, Haifa, 32000, Israel}
\author{I. Silber}
\email{itaisilber@tauex.tau.ac.il}
\affiliation{School of Physics and Astronomy, Tel Aviv University, Tel Aviv, 6997801, Israel}
\author{S. Sandik}
\affiliation{School of Physics and Astronomy, Tel Aviv University, Tel Aviv, 6997801, Israel}
\author{M. Lotem}
\affiliation{School of Physics and Astronomy, Tel Aviv University, Tel Aviv, 6997801, Israel}
\author{A. Ribak}
\affiliation{Department of Physics, Technion -- Israel Institute of Technology, Haifa, 32000, Israel}
\author{Y. Nitzav}
\affiliation{Department of Physics, Technion -- Israel Institute of Technology, Haifa, 32000, Israel}
\author{A.Yu.~Kuntsevich}
\affiliation{P.N. Lebedev Physical institute of the RAS, Moscow, 119991, Russia}
\author{O.A.~Sobolevskiy}
\affiliation{P.N. Lebedev Physical institute of the RAS, Moscow, 119991, Russia}
\author{Yu.G.~Selivanov}
\affiliation{P.N. Lebedev Physical institute of the RAS, Moscow, 119991, Russia}
\author{V.A.~Prudkoglyad}
\affiliation{P.N. Lebedev Physical institute of the RAS, Moscow, 119991, Russia}
\author{M.Shi}
\affiliation{Swiss Light Source, Paul Scherrer Institut, CH-5232 Villigen, Switzerland}
\author{L. Petaccia}
\affiliation{Elettra Sincrotrone Trieste, Strada Statale 14 km 163.5, 34149 Trieste, Italy}
\author{M. Goldstein}
\affiliation{School of Physics and Astronomy, Tel Aviv University, Tel Aviv, 6997801, Israel}
\author{Y. Dagan}
\email{yodagan@tauex.tau.ac.il}
\affiliation{School of Physics and Astronomy, Tel Aviv University, Tel Aviv, 6997801, Israel}
\author{A. Kanigel}
\email{amitk@physics.technion.ac.il}
\affiliation{Department of Physics, Technion -- Israel Institute of Technology, Haifa, 32000, Israel}

\begin{abstract}
Topological superconductivity is an exotic phase of matter in which the fully gapped superconducting bulk hosts gapless Majorana surface states protected by topology. Intercalation of copper, strontium or niobium between the quintuple layers of the topological insulator Bi$_2$Se$_3$ increases the carrier density and leads to superconductivity that is suggested to be topological. Here we study the electronic structure of strontium-intercalated Bi$_2$Se$_3$ using angle resolved photoemission spectroscopy (ARPES) and Shubnikov-de Haas (SdH) oscillations. Despite the apparent low Hall number of $\sim2 \times 10 ^{19}$cm$^{-3}$, we show that the Fermi surface has a shape of an open cylinder with a larger carrier density of $\sim 10 ^{20}$cm$^{-3}$. We suggest that superconductivity in intercalated Bi$_2$Se$_3$ emerges with the appearance of a quasi-2D open Fermi surface. 
\end{abstract}

\maketitle
A promising avenue towards achieving topological superconductivity is to induce superconductivity in materials whose electronic structure is topologically non-trivial. Intercalated \BiSe\ has been suggested as a possible candidate for a topological superconductor based on the appearance of a zero-bias feature in the tunnelling conductance \cite{Sasaki2011,Kirzhner2012,Levy2013,Tao2018} and nematicity in the superconducting state \cite{Matano2016,Kuntsevich2018,Kuntsevich2019, Yonezawa2016, Willa2018, Sun2019, Asaba2017, Tao2018}. A possible explanation to the zero bias peak is the existence of Majorana modes \cite{Sasaki2011,Levy2013,Tao2018}. Also, nematicity can be accounted for by a  two-component Ginzburg-Landau theory with nematic solutions corresponding to topological-superconductivity with odd-parity $E_u$ symmetry (the order parameter is odd under the operation $\mathrm{x\rightarrow-x}$ and $\mathrm{y\rightarrow-y}$). Such nematic solutions require an open, quasi-2D,  Fermi-Surface \cite{Fu2014,Wan2014,Venderbos2016,Roising2018}.

A simple criterion has been put forth for determining whether superconductivity in Bi$_2$Se$_3$-derived systems is 3D topological time-reversal-invariant (TRI): the Fermi surface (FS) has to enclose an odd number of time-reversal-invariant-momenta (TRIM) points \cite{Fu2010}. Superconducting A$_x$Bi$_2$Se$_3$ compounds are slightly electron-doped ($n\sim10^{19} -10^{20}$ cm$^{-3}$) with the bottom of the conduction band located at the center of the Brillouin zone \cite{Zhang2010}. In  Cu$_x$Bi$_2$Se$_3$, ARPES, Shubnikov-de Hass (SdH) and de Haas-van Alphen \cite{Lahoud2013,Lawson2014} experiments found an open FS that encloses two TRIM points ($\Gamma$ and Z) thus not fulfilling the aforementioned criterion but suggesting the possibility of weak 2D topological superconductivity. 

Increasing the carrier density of \BiSe\ via Cu intercalation results in a quick expansion of the FS towards the Z point and eventually a Lifshitz transition into an open cylindrical-like FS \cite{Lahoud2013}. Interestingly, Sr intercalated Bi$_2$Se$_3$ is reported to have an order of magnitude smaller carrier concentration as compared to the Cu intercalation, while still exhibiting superconductivity with similar critical temperature T$_c$ \cite{Shruti2015,Kuntsevich2018,Kuntsevich2019b}. This raises two questions: 1) Does \SrBiSe have a closed Fermi surface that allows for a 3D topological TRI superconductivity? 2) How can such a small amount of carriers produce a T$_c\simeq$3K? 

\begin{figure}[htbp!]
\centering
\includegraphics[width=1\linewidth]{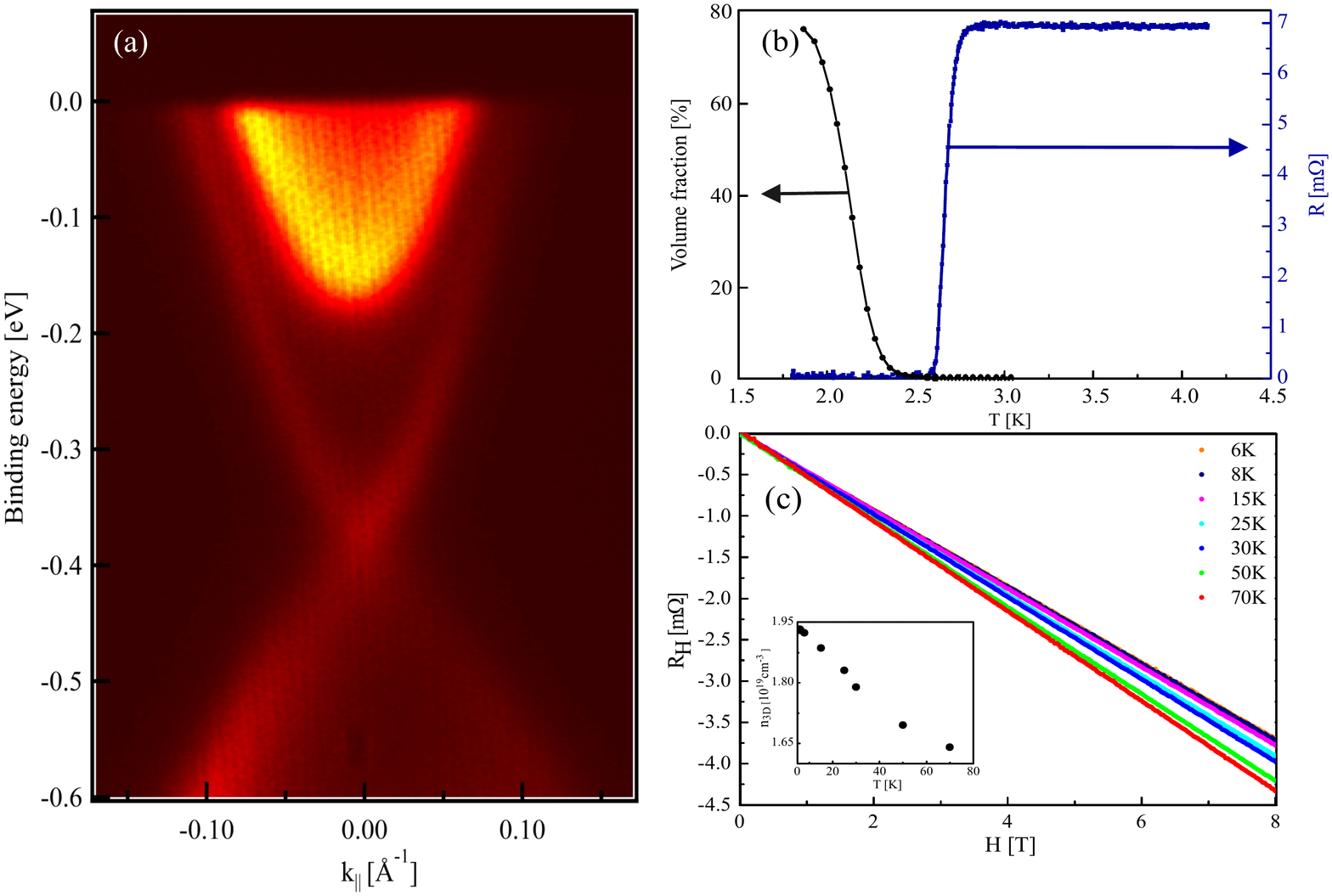}
\caption{Bulk superconductivity coexists with topological surface states in Sr$_x$Bi$_2$Se$_3$. (a) Typical ARPES spectra, taken using a 20~eV photon energy at $T=28.5$~K, displays the electron-like band and the surface states. The Dirac point lies about 375~meV below the Fermi-level. (b) Resistance and magnetization measurements of the sample measured in the Shubnikov-de-Hass part. The sharp resistive transition at 2.7K, along with the large superconducting volume fraction, clearly demonstrate a 3D bulk superconductivity. (c) Hall measurements at various temperatures. The linearity of the Hall coefficient with respect to the magnetic field is visible, while a clear temperature dependence is apparent. Inset: The carrier density inferred from the Hall measurements vs. temperature.}
\label{Fig1}
\end{figure}

In this letter we combine the SdH effect and ARPES measurements to determine the shape of the Fermi-surface, motivated by the relatively small Hall number of $\mathrm{Sr_{x}Bi_{2}Se_{3}}$. We study \SrBiSe samples with $\mathrm{x_{nominal}}=0.15$, crystal growth and chracterization are described in Ref.~\cite{Kuntsevich2018,Volosheniuk2019} and in the supplementary. The samples display Dirac-like surface states as well as a parabolic bulk band. A naive estimate of the carrier density from the Hall slope yields a carrier concentration of $\sim2 \times 10 ^{19}$cm$^{-3}$. This estimate is in good agreement with the binding-energy of the Dirac-point \cite{Lahoud2013}, but has a noticeable temperature dependence (see Fig.\ref{Fig1}). Bulk superconductivity is evident from the large SC volume fraction, which extrapolates to 100\% at zero temperature, with T$_c =$ 2.7K.

\par
\begin{figure}[htbp]
	
	\includegraphics[width=1\columnwidth]{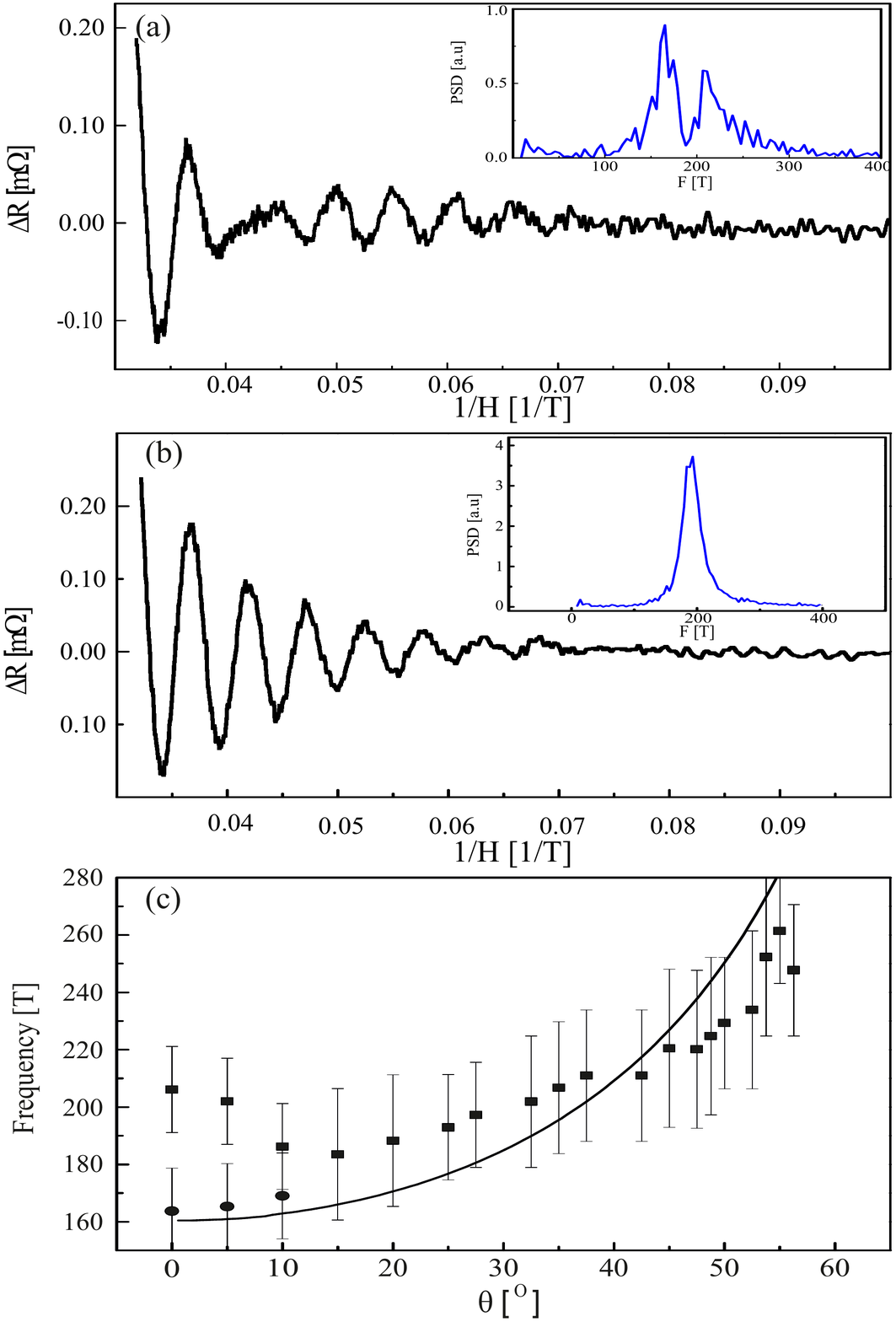}
	\caption{Angular dependence of the SdH frequency at 0.4 K. (a) Two frequencies are observed for $\theta \simeq 0^{\circ}$. The two frequencies point to an open Fermi surface as explained in the text. (b) As we rotate the sample to larger out of plane angles, we observe only one frequency above a certain angle, as can be seen, for example, at $\theta \simeq 20^{\circ}$. (c) SdH frequency as a function of angle. The solid line represent the expected angular dependence of one SdH frequency for an open cylinder. The insets in (a) and (b) are the Fourier power spectrum of the quantum oscillations vs inverse magnetic field $\mathrm{(1/H)}$. $\theta$ is the angle between the magnetic field direction and the c-axis.}
	\label{SdH}
	
\end{figure}

The frequency of the resistance-oscillations as a function of inverse magnetic field corresponds to the cross-sectional areas of the extrema of the FS perpendicular to the magnetic field \cite{Onsager1952}. We therefore rotate the magnetic field with respect to the c-axis ($\theta$) to resolve the shape of the FS. Our main finding using this method is the clear observation of two frequencies when the magnetic field is parallel to the c-axis, $\theta=0$ (see Fig.~\ref{SdH}(a)). 
 
This finding is a strong evidence for an open FS, as 
two frequencies will only show up if the FS have two extrema with different cross-sectional area. Therefore, A closed Fermi-surface would have two frequencies only for a complicated shape, which is inconsistent with previous ARPES measurements and theoretical DFT and $\boldmath{k\cdot p}$ calculations \cite{Zhang2010,Hashimoto2014}.

 The two frequencies persist over a wide temperature range. However, as we increase $\theta$ only one frequency remains visible in the SdH data beyond 10$^{\circ}$. The disappearance of the second frequency is probably due to a strong damping of the oscillations, possibly a consequence of different effective masses. The remaining frequency increases when rotating the field away from the c-axis, becoming unresolved for $\theta>$57$^{\circ}$. This frequency is well described by $1/\cos(\theta)$, as expected for an open cylindrical-type FS (see Fig.~ \ref{SdH}).

\par
 Due to the larger uncertainties for $\theta>50^{\circ}$, we found it necessary to support our analysis suggesting an open FS in $\mathrm{Sr_{x}Bi_{2}Se_{3}}$. We turn to ARPES measurements in which we vary the photon energy and thus scan the entire Brillouin zone at different k$_{z}$ \cite{Xu2011,Lahoud2013,Hufner2013}. The inner-potential that determines the conversion of photon-energy to momentum normal-to-the-surface is derived in the supplementary material.  

\begin{figure*}[t]
	
	\includegraphics[trim={0cm 0 0cm 0},clip,width=1\textwidth]{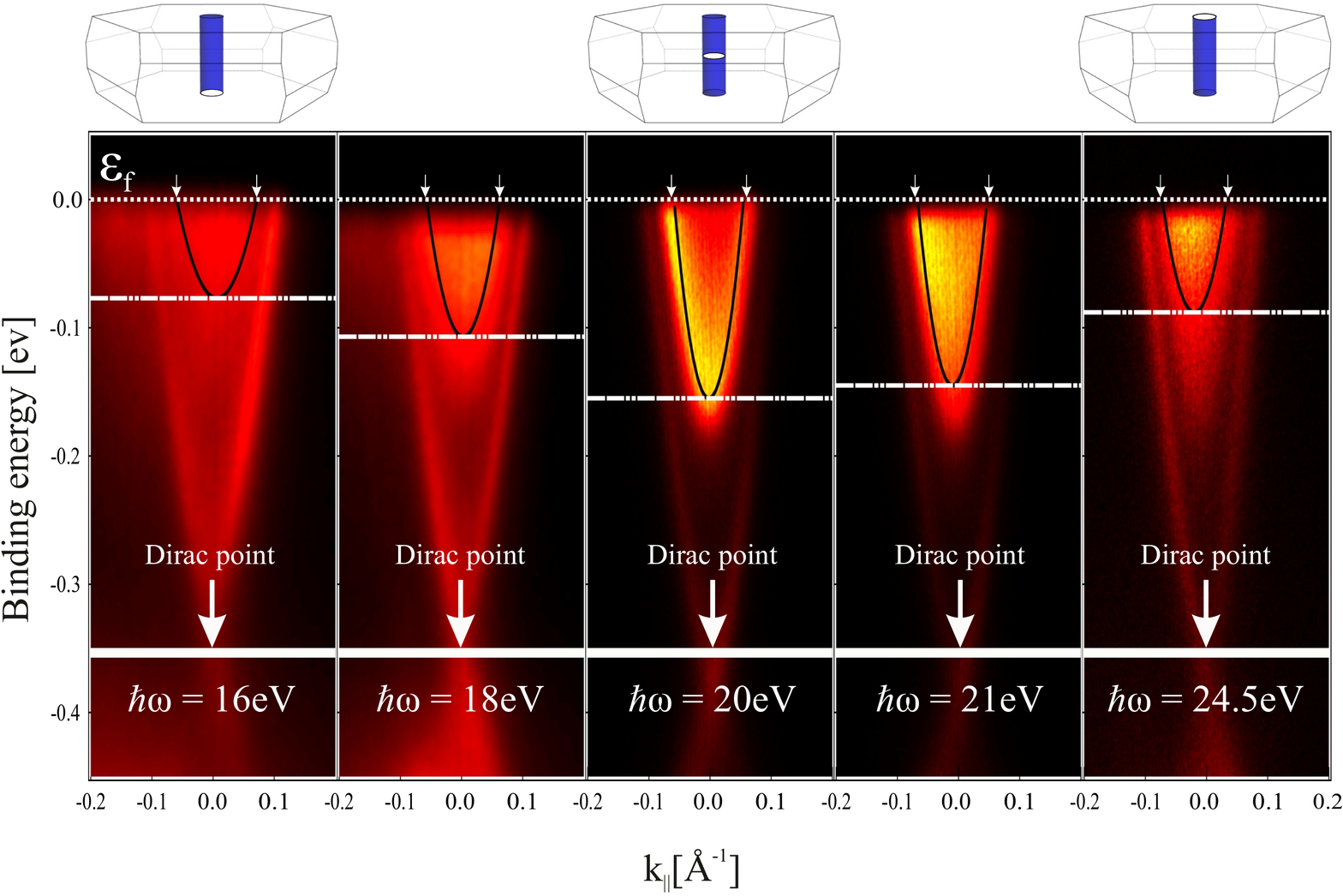}
	\caption{Band structure of $\mathrm{Sr_{x}Bi_{2}Se_{3}}$ samples measured at various photon energies, each photon energy corresponds to a different k$_z$ value (see supplementary material).  The two outmost panels are measurements at the two $\mathrm{Z}$ points, while the middle panel is a measurement at k$_z$=0. The 16 and 18 eV data were measured at the BaDElPh beam line at Elettra, and the 20, 21, 24.5 eV data were measured at the SIS beamline at the SLS,PSI. The small arrows mark the Fermi momentum, and the large arrow marks the Dirac point. $\epsilon_{F}$ and the minimum of the band are marked by dashed and dashed-dotted lines, respectively. In all panels the energy of the Dirac point is the same, as expected. Drawings on top: An illustration of the constant $k_z$ plane at which the spectra in the panel below it was measured.}
	\label{ARPES}

\end{figure*}

Each panel in Fig. \ref{ARPES} displays the dispersion of the conduction band with respect to the momentum parallel to the surface for different photon energies. 
We find intensity corresponding to the bulk band for all measured  k$_z$ value, including the $\Gamma$ and Z points. Our data showing finite density of states at both Z-points support our conclusion of open FS. For comparison, we note that no intensity is seen by ARPES at the k$_z$ corresponding to the Z points for doped \BiSe~ having a closed FS \cite{Lahoud2013,Tanaka2012}. The bandwidth changes with k$_z$ and is maximal at the $\Gamma$-point, similarly to the case of \CuBiSe and consistent with our SdH data. Furthermore, the  Fermi-momentum is independent of k$_z$, suggesting a nearly perfect cylinder. The constant Fermi-momentum in the $\mathrm{k_x-k_y}$ plane for different $k_z$ suggests that the effective mass of the band depends on $k_z$ (see supplementary data).   
\par
The temperature and intercalation dependence of the Hall coefficient casts a doubt whether it can be used for an accurate determination of the carrier density. The Hall number is almost independent of Sr intercalation, albeit this intercalation is the main mechanism for doping mobile electrons in the sample\cite{Liu2015,Volosheniuk2019}. Together with the unexpected large temperature dependence of the Hall number we conclude that the Hall-inferred density is unreliable. Doped Bi$_2$Se$_3$ has a single parabolic band close to the Fermi-energy, and if we assume that the SdH signal originates only from the conducting bulk, the interpretation of the data is straightforward and gives an exact information about the volume of the Fermi-surface and the carrier density. We therefore revisit our previous results and calculate the carrier density from the volume in momentum space as inferred from ARPES and SdH. The Fermi momentum from the ARPES, which is a direct method, gives $n=6.5\pm1\times10^{19}$~cm$^{-3}$ for a non corrugated cylinder. The SdH numbers for this sample and for other samples are summarized in Table \ref{ta}.

\par
A similar Fermi-surface topology is found in {\it all} intercalated Bi$_2$Se$_3$ samples exhibiting superconductivity\cite{Lahoud2013,Lawson2016}. This observation raises the question of why does superconductivity appear concomitantly with an open Fermi-surface?

\begin{figure}[htbp]
	
	\includegraphics[width=1\linewidth]{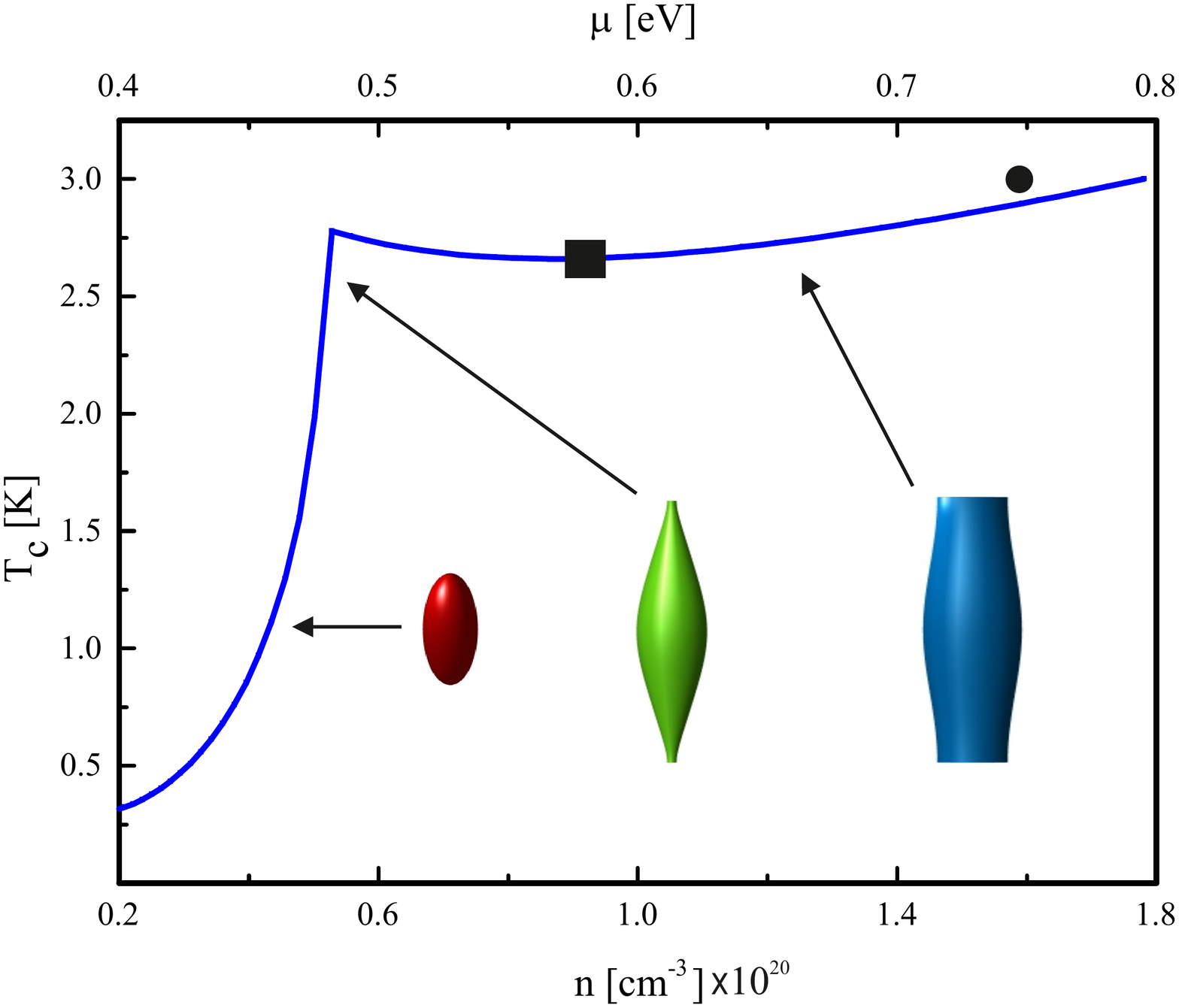}
	\caption{Theoretical chemical potential/carrier density dependence of the superconducting critical temperature, obtained by combining DFT-based band structure with the BCS theory. Note the chemical potential calculated is about 3 times larger compared to the maximum bandwidth obtaines by ARPES. Two experimental points are presented, a black dot for \CuBiSe\cite{Lahoud2013} and a black square for \SrBiSe, showing the consistency of the theoretical model.}
	\label{Model}
	
\end{figure}

Intuitively, one might expect that the opening of the FS will result in a quasi-2D band structure that has a larger density of states compared with the 3D case, for same carrier density. In the framework of weak-coupling BCS theory, the increase in density of states would enhance the critical temperature T$_c$. We use a simple model to quantify this intuitive idea, using weak-coupling limit BCS theory and the band-structure of $\mathrm{Bi_{2}Se_{3}}$, based on DFT calculations for pure $\mathrm{Bi_{2}Se_{3}}$ \cite{Zhang2009,Sasaki2011}. To account for the effect of the intercalation 
the coupling amplitudes in the $\hat{z}$ direction are varied as a function of the chemical potential \cite{Hashimoto2014}. The FS of this band-structure is a narrow ellipsoid \textit{closed} surface at low carrier densities and a cylindrical \textit{open} surface at high carrier densities, as shown in Fig.~\ref{Model}, and in agreement with the experimental observation. 
\par
We can gauge the effect of the Lifshitz transition by plugging the density of states $g\left(\mu\right)$ into the standard weak coupling BCS \cite{Cooper1957} expression for the critical temperature $T_{c}$ in the weak coupling regime:

\begin{equation}
k_{B}T_{c}\approx1.14\hbar\omega_{D}\exp\left(-\frac{1}{V\cdot g\left(\mu\right)}\right),\label{eq:BCS-Tc}
\end{equation}
where $\omega_{D}$ is the Debye frequency and $V$ is the electron-phonon coupling constant. We use the known $\omega_{D}$ and adjust $V$ to match the critical temperature of the system at large enough carrier densities where $g\left(\mu\right)$ is approximately constant. Assuming that $V$ is independent of carrier density, we can plug it back into Eq.~(\ref{eq:BCS-Tc}) and estimate $T_{c}$ for different carrier densities. As shown in Fig.~\ref{Model}, the critical temperature of the open FS ($n>5\times10^{19}$cm$^{-3}$) is approximately constant, but drops sharply at carrier densities below the Lifshitz transition once the Fermi surface closes. Two experimental points $T_c(n)$ are indicated on the plot, matching the critical temperatures and carrier densities for $\mathrm{Cu_xBi_2Se_3}$ and $\mathrm{Sr_{x}Bi_{2}Se_{3}}$ deduced from SdH measurements.

\begin{table}[htbp!]
	\centering 
	\begin{tabular}{|c|c|c|c|}
		\hline 
		Sample & $n_{SdH}$~cm$^{-3}$&  $n_{Hall}$~cm$^{-3}$& ref. \\  \hline 
		\BiSe\ (not SC) & 1.9$\times10^{18}$                 &$3.7\times10^{17} $               & \cite{Lahoud2013} \\ \hline 
		\BiSe\ (not SC) & 1.8$\times10^{19}  $               &$2.3\times10^{19} $               &\cite{Lahoud2013}  \\ \hline 
		$\mathrm{Cu_x}$\BiSe\ (SC) & $1.6\times10^{20}$                 &$2.7\times10^{20} $               &\cite{Lahoud2013}  \\ \hline 
		$\mathrm{Sr_x}$\BiSe\ (SC) & $9.3\times10^{19} $                &$2\times10^{19*} $               & This paper \\ \hline 
	\end{tabular} 
	\caption{Comparison between the Hall carrier density and the SdH carrier density. * - The Hall density of $\mathrm{Sr_{x}Bi_{2}Se_{3}}$ is taken at $T=6$~K.}
	\label{ta}
\end{table}

The main finding is that superconducting samples always have large ($\geq6.5\times10^{19}$cm$^{-3}$) carrier density, which results in an open FS, consistent with our simplified model and measurements done using different techniques.
From the theoretical point of view\cite{Fu2010}, a 3D topological superconductor is not achievable with our suggested Fermi-surface. This still leaves the door open for other scenarios of weak 2D topological superconductivity \cite{Fu2008}.

An open Fermi-surface is consistent with two more observations. First, the most realistic physical mechanism of superconducting pairing in \SrBiSe is the electron-phonon interaction, that occurs at small phonon momenta \cite{Wan2014,Brydon2014}. An open Fermi-surface allows this mechanism to be energetically favourable\cite{Wan2014}, as confirmed recently by inelastic neutron scattering \cite{Wang2019}.  Second, as shown in Ref.\cite{Hao2019}, for an odd-parity E$_u$ pairing, the nodeless superconducting gap observed in tunnelling measurements in Cu$_x$Bi$_2$Se$_3$\cite{Sasaki2011,Levy2013,Tao2018} and Sr$_x$Bi$_2$Se \cite{Du2017} is consistent only with an open Fermi-surface. 

In addition, the suggested mean-field picture with the very weak dependence of T$_c$ on the density is  in agreement with the observation of a Hall density independent $T_c$ in Cu-co-doped Sr$_x$Bi$_2$Se$_3$ crystals \cite{Kuntsevich2019b}.

Finally, we note that according to our simplified model a small region in the carrier density (4-5$\times10^{19}$~cm$^3$) - temperature phase diagram may allow for a 3D topological superconductivity. This would require precise tuning of carrier density, possibly using gate voltage or proper co-doping. 

\section{Acknowledgments}
A. Almoalem and I. Silber contributed equally to this work. ARPES experiments were conducted at
the Surface/Interface Spectroscopy (SIS) beamline of the Swiss Light Source at the Paul Scherrer Institut in Villigen, Switzerland and at the BaDElPh beamline of Elettra. The high magnetic field measurements were performed at the National High Magnetic Field Laboratory, which is funded by the National Science Foundation through DMR-1157490 and the U.S. Department of Energy and the State of Florida. The work on the Russian site (crystal growth and magnetotransport measurements) was supported by the Russian Science Foundation (Grant No. 17-12-01544). Magnetotransport measurements were performed using the equipment of the LPI shared facility center. Work at Tel Aviv University and at the Technion is supported by the Israeli Science Foundation under grant number 382/17 and 320/17 respectively. Theoretical work is supported by Israel Science Foundation (Grant No. 227/15), US-Israel Binational Science Foundation (Grant No. 2016224). We acknowledge useful discussions with Assa Auerbach.

\bibliographystyle{apsrev4-1}
\bibliography{SrBiSebib}
\end{document}